# An Intelligent System For Effective Forest Fire Detection Using Spatial Data

K.Angayarkkani
Senior lecturer
Department of ComputerApplications
D.G. Vaishnav College, Arumbakkam, Chennai
.

Dr.N.Radhakrishnan
Geocare Research Foundation
#23/30, First main Road,
Pammal, Chennai - 600 075, India
.

*Abstract*— The explosive growth of spatial data and extensive utilization of spatial databases emphasize the necessity for the automated discovery of spatial knowledge. In modern times, spatial data mining has emerged as an area of voluminous research. Forest fires are a chief environmental concern, causing economical and ecological damage while endangering human lives across the world. The fast or early detection of forest fires is a vital element for controlling such phenomenon. The application of remote sensing is at present a significant method for forest fires monitoring, particularly in vast and remote areas. Different methods have been presented by researchers for forest fire detection. The motivation behind this research is to obtain beneficial information from images in the forest spatial data and use the same in the determination of regions at the risk of fires by utilizing Image Processing and Artificial Intelligence techniques. This paper presents an intelligent system to detect the presence of forest fires in the forest spatial data using Artificial Neural Networks. The digital images in the forest spatial data are converted from RGB to XYZ color space and then segmented by employing anisotropic diffusion to identify the fire regions. Subsequently, Radial Basis Function Neural Network is employed in the design of the intelligent system, which is trained with the color space values of the segmented fire regions. Extensive experimental assessments on publicly available spatial data illustrated the efficiency of the proposed system in effectively detecting forest fires.

## I. INTRODUCTION

The rapid progress in scientific data collection has led to enormous and ever-increasing quantity of data making it unfeasible to be manually interpreted. Therefore, the development of novel techniques and tools in assist for humans, aiding in the transformation of data into useful knowledge, has been the heart of the comparatively new and interdisciplinary research area called the "Knowledge Discovery in Databases (KDD)" [3]. Data mining is the vital step in KDD, which facilitates the discovery of buried but valuable knowledge from enormous databases. Data Mining is formally defined as "The non-trivial extraction of inherent, new, and potentially valuable information from databases" [5].Data mining combines machine learning, pattern recognition, statistics, databases, and visualization techniques into a single unit so as to enhance efficient information extraction from large databases [6]. Data mining techniques profit a number of fields like marketing, manufacturing, process control, fraud detection and network management. Other than this, a huge variety of data sets like market basket data, web data, DNA data, text data, and spatial data [7] have benefited as well.

The progress in scientific data collection has resulted in huge and continuously rising amount of spatial data [1]. Thus the need, for automated discovery of spatial knowledge from massive amount of spatial data, arises. The process of identifying previously hidden but valuable information from vast spatial databases is known as spatial data mining. It is comparatively tedious to extract patterns of value and interest from the spatial databases owing to the complexity of spatial data types, spatial relationships, and spatial autocorrelation than that of the conventional numeric and categorical data [2]. Spatial data mining technologies facilitate the comprehension of spatial data, discovery of relationships among spatial and non-spatial variables, determination of the spatial distribution patterns of a specific phenomenon further supporting the envisagement of the pattern trends. The elemental parts of spatial data mining are spatial statistics and data mining. Spatial data mining techniques involves visual interpretation and analysis, spatial and attribute query and selection, characterization, generalization and classification, detection of spatial and non spatial association rules, clustering analysis and spatial regression in addition to a wide variety of other fields [8].

Spatial data mining and relational data mining vary from one another because of the fact that in the former the attributes of the neighbors of some object of interest need to be taken into consideration as well, since they have a prominent influence on the object [9]. Some distinguishing characteristics of spatial data that forbid the usage of regular data mining algorithms include: (i) rich data types (e.g., extended spatial objects) (ii) inherent spatial relationships between the variables, (iii) other factors that influence the observations and (iv) spatial autocorrelation among the characteristics [2]. The extraction of patterns of interest and rules from the spatial data sets like the remotely sensed imagery and related ground data significantly benefits the application areas like precision agriculture, community planning, resource discovery and more[10]. Spatial data mining is comprehensively employed in change detection, modeling deforestation, disaster analysis, forest fire detection and other related fields. Our research focuses on the detection of forest fires from the spatial data analogous to forest regions.

*A. Forest Fires*

For a long time, fires have been a source of trouble. Fires have notable influence over the ecological and economic utilities of the forest, being a prime constituent in a great number of forest ecosystems [11]. Past has witnessed multiple instances of forest and wild land fires. Fires play a remarkable role in determining landscape structure, pattern and eventually the species composition of ecosystems. The integral part of the ecological role of the forest fires [13] is formed by the controlling factors like the plant community development, soil nutrient availability and biological diversity. Fires are considered as a significant environmental issue because they cause prominent economical and ecological damage despite endangering the human lives [12]. Due to the forest fires, several hundred million hectares (ha) of forest and other vegetation are destroyed every year [14].

Occasionally, forest fires have forced the evacuation of susceptible communities in addition to heavy damages amounting to millions of dollars. As per the forest Survey of India 19.27% or 63.3 million ha of the Indian land has been classified as forest area, of which 38 million ha alone are hoarded with resources in great quantity (crown density above 40%).Thus the country's forests face a huge threat. Degradation caused by forest fires [15] jeopardizes the Indian forests. Fires caused huge damage in the year 2007 affecting huge territories in addition to the prominent number of human casualties [16]. Forest fires remains to be a potential threat to ecological systems, infrastructure and human lives. The practical and effective option to minimize the damage caused by the forest fire is to detect the fires at their early stages and reacting fast to prevent the spread of the fire. Hereafter, hefty efforts have been taken to ease the early detection of forest fires, usually being carried out with the help of human surveillance. Forest fire, Drought, Flood and many other phenomena especially the ones with large spatial extent are some of the spatial phenomena which contain predictable spatial patterns that are evident through remote sensing Images/products.

*B. Our Contributions*

In our earlier work [45], we have presented an efficient forest fire detection system using Fuzzy logic. The primary intention of this research is to extract valuable information from spatial data and employ them for locating the regions vulnerable to forest fire with the aid of Image Processing and Artificial Intelligence techniques. This paper presents an intelligent system that is capable of detecting forest fires. The presented intelligent system utilizes the images in the spatial data that corresponds to forest regions, obtained from remote sensing. The Radial Basis Function Neural Network is employed in the design of the presented intelligent system. The images in the forest spatial data with the presence of fires are utilized in training the neural network. Initially, the digital images in the forest spatial data are converted from RGB to $XYZ$ color space. Then, the segmentation of the image in $XYZ$ color space is carried out with the aid of the renowned anisotropic diffusion approach. The $XYZ$ color space values of the regions with fires, resulting from the segmentation, are fed as input to training the radial basis function neural network. For a given $XYZ$ color space value of a pixel, the trained radial basis function neural network will identify whether that pixel corresponds to fire region or not. The presented intelligent system effectively detects forest fire, which is very well illustrated by the experimental evaluation on the publicly available spatial data.

The rest of the paper is organized as follows: Section II presents a brief review of some recent researches existing in the literature related to forest fire detection. A concise description of the concepts utilized in the presented intelligent system is given in Section III. The proposed intelligent system for effective forest fire detection is presented in Section IV. The experimental results are given in Section V. The conclusions are summed up in Section VI.

II. REVIEW OF RELATED RESEARCHES

The proposed research has been motivated by several earlier researches in the literature related to forest fire detection using spatial data and artificial intelligence techniques. A concise description of some of the recent researches is given in this section.

Armando et al. [17] have studied on the automatic recognition of smoke signatures in lidar signals attained from very small-scale experimental forest fires using neural-network algorithms. A scheme of multi-sensorial integrated systems for early detection of forest fires has been presented by Ollero et al. [18]. The system presented by the authors uses infrared images, visual images, and data from sensors, maps and models. To facilitate the minimization of perception errors and the improvement in reliability of the detection process, it is necessary for the integration of sensors, territory knowledge and expertise, according to their study.

An improved fire detection algorithm which provides increased sensitivity to smaller, cooler fires as well as a significantly lower false alarm rate has been presented by Louis Giglio et al. [19]. The Theoretical simulation and high-resolution Advanced Space borne Thermal Emission and Reflection Radiometer (ASTER) scenes are employed to establish the performance of their algorithm. Seng Chuan Tay et al. [20] have presented an approach to reduce the false alarms in the hotspots of forest fire regions which uses geographical coordinates of hot spots in forest fire regions for detection of likely fire points. The authors employ clustering and Hough transformation to determine regular patterns in the derived hotspots and classify them as false alarms on the assumption that fires generally do not spread in regular patterns such as straight lines. In this work demonstrate the application of spatial data mining for the reduction of false alarm from the set of hot spots is derived from NOAA images.

A graph based forest fire detection algorithm based on spatial outlier detection methods has been presented by Young Gi Byun et al. [21]. By using the spatial statistics the authors have achieved spatial variation in their algorithm. This algorithm illustrates higher user and producer accuracies, when compared with the MODIS fire product provided by the NASA MODIS science team. The ordinary scatter plot algorithm was proved to be inefficient by the authors because it is insensitive to small fires, while Moran's scatter plot was also weak because of the numerical criterion's absence for spatial variation which requires a more and less high commission error.

An approach to predict forest fires in Slovenia using different data mining techniques has been presented by Daniela Stojanova et al. [22]. The authors have employed the predictive models based on the data from a GIS (Geographical Information System) and the weather prediction model - Aladin and MODIS satellite data. The work examined three different datasets: one for the Kras region, one for Primorska region and one for continental Slovenia. The researchers demonstrated that Bagging and boosting of decision trees offers the best results in terms of accuracy for all three datasets. Yasar Guneri Sahin [23] has proposed a mobile biological sensor system for prior detection of forest fires which utilizes animals as mobile biological sensors. This system is based on the existing animals tracking systems used for the zoological studies. The work illustrates that the combination of these fields may lead to instantaneous development of animal tracking as well as forest fire detection. A number of serious forest fires were detected by the system in the earliest, which reduced their effect and therefore contributes to the reduction of the speed of global warming.

A fully automated method of forest fire detection from TIR satellite images on the basis of random field theory has been presented by Florent Lafarge et al. [24]. The results of the system rely only on the confidence coefficient. The obtained values for the both detection rate and false alarm rate were convincing. The estimation of fire propagation direction presents interesting information associated to the evolution of the fires. In Movaghati et al. [25], the capability of agents to be applied in processing of remote sensing imagery has been studied. An agent based approach for forest fire detection has been presented in this paper. The tests used in MODIS version 4 contextual fire detection algorithms were used by the agents to determine agent behavioral responses. The performance of their algorithm was compared against that of MODIS version 4 contextual fire detection algorithm and ground-based measurements. The results portray a good agreement between the algorithms and field data.

In our earlier work [45], we have presented an efficient system to detect forest fires using spatial data collected from forest. Image Processing and Artificial Intelligence techniques were utilized in the design of the presented system. Anisotropic diffusion and fuzzy logic are employed for segmentation and fire detection processes respectively. The images are converted to YCbCr color space and segmentation is performed. The Cr value of YCbCr color space of fire pixels is utilized in the formation of fuzzy sets and fuzzy rules are derived from the formed fuzzy sets. The publicly available spatial data has been employed in the evaluation process. The fuzzy rules derived using the presented system, have successfully detected the forest fires in the spatial data.

III. DESCRIPTION OF CONCEPTS UTILIZED IN THE PRESENTED INTELLIGENT SYSTEM

The concepts utilized in the presented intelligent system for effective forest fire detection such as color space, anisotropic diffusion segmentation and artificial neural networks are detailed in this section.

*A. Color Space Conversion*

A color space is defined as a means by which the specification, creation and visualization of colors is performed. A computer screen produces colors based on the varied combinations of red, green and blue phosphor emission required to form a color. Typically color is represented by three coordinates or parameters [26]. The location of the color in the color space is exemplified by these parameters. Color space conversion is defined as the transformation and description of a color from one source to another. Normally, color space conversion is performed while converting an image that is represented in one color space to another color space, with the objective of making the translated image appear as similar as possible to the original. The commonly used color spaces are RGB, CIE $XYZ$, CIE YUV, CIE L*a*b*, YCbCr and HSV. In the proposed intelligent system, the images in RGB color space are converted to $XYZ$ color space.

*1) Cie Xyz Color Space:* CIE $XYZ$ color space [27] is one of the first mathematically defined color spaces created by the International Commission on Illumination in 1931. Any color can be generated as a mixture of three other colors or "Tristimuli" and commonly RGB for CRT based systems (TV, computer) or $XYZ$ (fundamental measurements). The $XYZ$ color space is defined such that all visible colors can be represented using only positive values, and, the Y value is luminance. As a result, the colors of the $XYZ$ primaries themselves are invisible [28]. The chromaticity diagram is extremely non-linear, in that a vector of unit magnitude denoting the difference between two chromaticities is not uniformly visible. A 3x3 matrix transform is used to transform the $RGB$ values in a particular set of primaries to and from CIE $XYZ$. These transformations involve tristimulus values which are a set of three linear-light components that conform to the CIE color-matching functions. CIE $XYZ$ is a special set of tristimulus values. The equations to convert $RGB$ into $XYZ$ color space are as follows:

$$\begin{bmatrix} X \\ Y \\ Z \end{bmatrix} = \begin{bmatrix} 0.412453 & 0.357580 & 0.180423 \\ 0.212671 & 0.715160 & 0.72169 \\ 0.019334 & 0.119193 & 0.950227 \end{bmatrix} * \begin{bmatrix} R \\ G \\ B \end{bmatrix} \quad (1)$$

*B. Anisotropic Diffusion Segmentation*

Image segmentation is defined as a low-level image processing task meant for partitioning an image into identical regions [29]. The segmentation results can be used to identify the regions of interest and objects in the scene that is very beneficial to the subsequent image analysis. Because of the fact that the inherent multi-features not only possess non linear relation independently but also encompass inter-feature dependency between R, G, and B, color image segmentation is more monotonous when compared to the grey image segmentation. In our system, we have employed an anisotropic diffusion approach for the segmentation of images in the spatial data. The segmentation is carried out on the $XYZ$ color space converted image.

Literature offers numerous models of linear and nonlinear diffusion for achieving image smoothing and segmentation. Nonlinear anisotropic diffusion has been one of the commonly used approaches by researchers [30], [31] in their works. The anisotropic diffusion enhances the response of edge detection algorithms by a series of operations namely: smoothing the image interiors to emphasize boundaries for segmentation, eliminating the spurious detail and eradicating noise from images efficiently [34]. The relaxation processes that implement anisotropic diffusion tends to leave out the low frequency objects that are complex to be dispersed without over-processing the image.

Anisotropic diffusion in image processing discretizes the family of continuous partial differential equations, which incorporate both the physical processes of diffusion and the Laplacian. Provided that there are no sinks or sources that exist [32], the following equation formulates the abovementioned process (for any dimension):

$$\frac{\partial}{\partial t} u(\overline{x},t) = div(c(\overline{x},t) \nabla u(\overline{x},t)) \quad (2)$$

Diffusion strength is controlled by $c(x,t)$. Vector $x$ represents the spatial coordinate(s). The ordering parameter is the variable $t$. The function $u(x,t)$ is considered as image intensity $I(x,t)$ [33].

*C. Artificial Neural Networks (ANN)*

Artificial Neural Networks are a branch of the artificial intelligence, developed to reproduce human reason and intelligence. ANN possesses the abilities to recognize patterns, manage data and learn like the brain [35]. The weights and the input-output function (transfer function) that is specified for the units are used to characterize the behavior of an ANN [37]. The most significant pros in using artificial neural networks are solving the very complex problems of conventional technologies, not formulating an algorithmic solution or using the very complex solution [35]. In the presented intelligent system, Radial Basis Function Neural Network is employed and is detailed below.

*1) Radial Basis Function Neural Network (RBFNN):* In the late 80's, a variant of artificial neural network emerged by the name, Radial Basis Functions. Nevertheless, their roots are well-established in much older pattern recognition techniques for instance potential functions, clustering, functional approximation, spline interpolation and mixture models [39]. Radial Basis Function Neural Network (RBFNN) is based on supervised learning. RBF networks were autonomously proposed by many researchers [40], [41], [42], [43], [44] and are a popular variant to the MultiLayer Perceptron MLP. RBF networks are also excellent at modeling non-linear data and can be trained in one stage rather than using an iterative process as in MLP and also learn the given application speedily. The RBF network has a feed forward structure consisting of a single hidden layer of $J$ locally tuned units, which are fully interconnected to an output layer of $L$ linear units. All hidden units concurrently receive the $n$-dimensional real valued input vector $X$ (Figure. 1). The prime difference from that of MLP is the absence of hidden-layer weights. The hidden-unit outputs are not computed using the weighted-sum mechanism/sigmoid activation; rather each hidden-unit output $Z_j$ is obtained by closeness of the input $X$ to an $n$-dimensional parameter vector $\mu_j$ associated with the $j^{th}$ hidden unit [4]. The response characteristics of the $j^{th}$ hidden unit ($j = 1,2,......,J$) is assumed as,

$$Z_j = K\left(\frac{\|X - \mu_j\|}{\sigma_j^2}\right) \quad (3)$$

Where $K$ is a strictly positive radially symmetric function (kernel) with a unique maximum at its 'centre' $\mu_j$ and which drops off rapidly to zero away from the centre. The parameter $\sigma_j$ is the width of the receptive field in the input space from unit $j$. This implies that $Z_j$ has an appreciable value only when the distance $\|X - \mu_j\|$ is smaller than the width $\sigma_j$. Given an input vector $X$, the output of the RBF network is the $L$-dimensional activity vector $Y$, whose $l^{th}$ component $(l = 1,2,....,L)$ is given by [36],

$$Y_l(X) = \sum_{j=1}^{J} w_{lj} Z_j(X) \quad (4)$$

For $l = 1$, mapping of (3) is similar to a polynomial threshold gate. However, in the RBF network, a choice is made to use radially symmetric kernels as 'hidden units'.

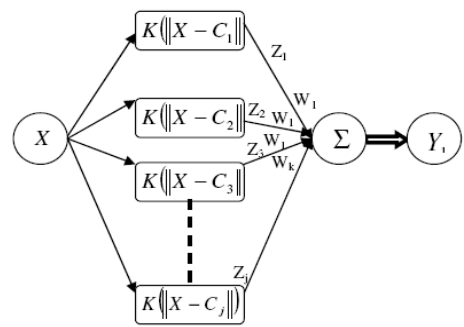

Figure1. Radial Basis Function Neural Network (RBFNN)

RBF networks are best suited for approximating continuous or piecewise continuous real-valued mapping $f: R^n \to R^L$, where n is sufficiently small. These approximation problems include classification problems as a special case. From (3) and (4), the RBF network can be viewed as approximating a desired function $f(X)$ by superposition of non-orthogonal, bell-shaped basis functions. The degree of accuracy of these RBF networks can be controlled by three parameters: the number of basis functions used, their location and their width [38].

IV. INTELLIGENT SYSTEM FOR EFFECTIVE FOREST FIRE DETECTION

The proposed intelligent system for effective detection of forest fires is presented in this section. The spatial data collected from forest regions are utilized by the presented intelligent system. With the aid of the images in the spatial data, forest fire detection is performed. The Radial Basis Function Neural Network is employed in the design of the presented intelligent system. The images in the forest spatial data with the presence of fires are employed in training the radial basis function neural network. Initially, the images with the presence of fires are converted from RGB to $XYZ$ color space. The color space conversion from RGB to XYZ is carried out with the help of (1). Figure 2 shows the image in RGB color space and its corresponding XYZ color space converted image.

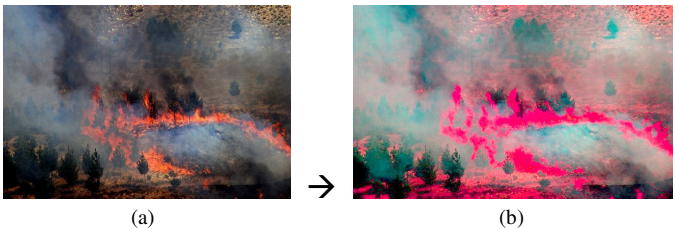

(a) $\rightarrow$ (b)

Figure 2. a) Image in RGB color space, b) XYZ color space converted image

Afterwards, the $XYZ$ color space converted image is segmented using anisotropic diffusion segmentation, which locates the regions of fire. The result of anisotropic diffusion segmentation is depicted in Figure 3.

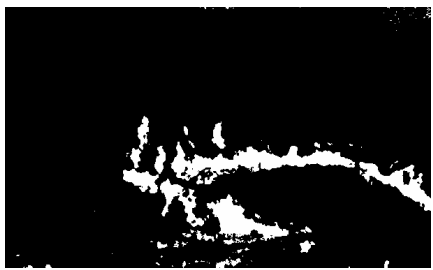

Figure 3. Anisotropic Diffusion Segmented image

The regions of fire obtained using the segmentation is utilized in training the radial basis function neural network. The radial basis function neural network is trained with the $XYZ$ color space values of the pixels that belong to fire regions. With the help of the trained neural network, we can effectively detect the presence of forest fires in an image. The presence of forest fire in an image is detected using the following steps. Initially the image is converted from RGB to $XYZ$ color space. Then, the color space converted image is segmented using anisotropic diffusion segmentation. Subsequently, the $XYZ$ color space values of pixels in the segmented regions are fed as input to the trained neural network for detecting the presence of fires. The designed intelligent system will aid the people in surveillance to detect forest fires and to take appropriate actions.

V. EXPERIMENTAL RESULTS

This section presents the results obtained from the experimentation on the presented intelligent system. The proposed intelligent system is implemented in MATLAB (Matlab 7.4). The publicly available forest spatial data with the presence of fires are employed in training the radial basis function neural network. Consequently, forest spatial data with and without the presence of fires are fed as input to the proposed system for evaluation. The presence of fires is detected effectively by the presented intelligent system with the aid of the trained neural network. The intermediate results of the presented system are depicted in Figure 4. From the results we can conclude that the presented intelligent system can be used for effectively detecting forest fires in the spatial data using artificial intelligence techniques.

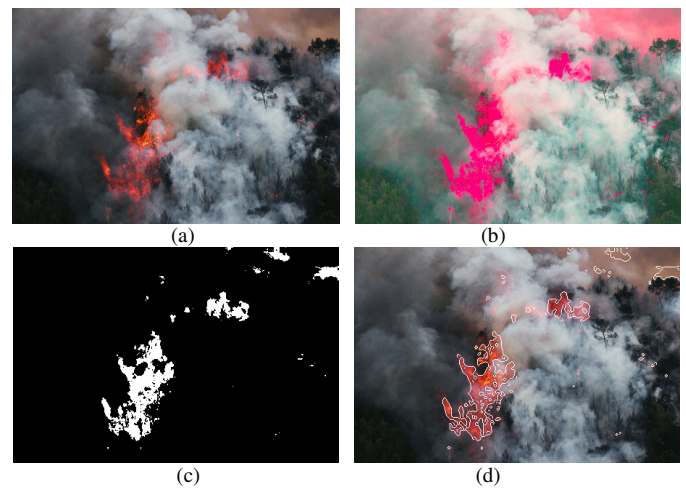

(a) (b)

(c) (d)

Figure. 4 Intermediate results of the presented intelligent system a) Input Image, b) XYZ color space converted image, c) Output of Anisotropic Diffusion Segmentation, d) Fire detected Image

VI. CONCLUSION

Forest fires cause noteworthy environmental demolition while menacing human lives. In the last two decades, a significant effort was made to develop automatic detection tools that could aid the Fire Management Systems (FMS). The three chief trends used for the detection of forest fires are: the use of satellite data, infrared/smoke scanners and local sensors (e.g. meteorological). In this paper, we have presented an intelligent system for effective forest fire detection using spatial data. The proposed system made use of image processing and artificial intelligence techniques. The images in the spatial data, obtained from remote sensing, have been utilized by the presented system for the detection of forest fires. The color space conversion, anisotropic diffusion segmentation and Radial Basis Function Neural Networks have been employed in the presented intelligent system. The experimental results have demonstrated the effectiveness of the proposed intelligent system in detecting forest fires using spatial data.

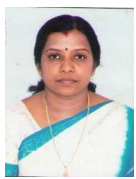

The author is a postgraduate in Computer Science followed by Master of Philosophy in Computer Science. The author has thirteen years of teaching experience in various fields of computer science. She has enrolled in Mother Teresa Women's University Kodaikanal for her Ph.D. doctoral degree. The author is currently doing research work on spatial data mining and image processing based techniques.

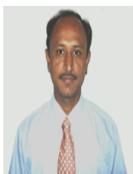

The co-author is a post-graduate in Applied Geology (1985) followed by **M.Tech** degree in **Remote Sensing** (1990) and have completed the **Ph.D doctoral degree** in 1999 on **spatial techniques -** Remote sensing, GPS and GIS - **watershed environment**. The co-author, to his credit, has **Eighteen years** of research and field **experience** in spatial data and geoinformatics - Remote sensing, GIS and GPS - applications. To his credit, he has published twelve research papers in refereed journals – national and international – and international conferences and two papers are under peer review. He has also had the distinct honor of acting as Chairperson for a session on "Ecosystem and Bio-diversity" in an International conference held at Tsukuba University, Ibaraki, Japan, apart from participating many national level seminars, workshops and training programs. He has also been involved in consultancy service to UNESCO, New Delhi, and developed Computer Based Learning tutorial on Geology using VB as front end tool with various graphic utilities (A/V Support) under ICT program. He has been appointed as Lesson writer for M.Sc Geoinformatics covering Satellite Remote sensing and GIS by Annamalai University and has been acting as Resource person for Academic Institutions.